\input{aipcheck}

\documentclass[
    ,final            
  ]
  {aipproc}

\layoutstyle{8x11single}

\usepackage{amsmath}
\usepackage{graphicx}

\newcommand{\etal}{\emph{et al.}}

\newcommand{\Od}{\mathcal{O}}
\newcommand{\muq}{\mu_q}

\begin{document}

\title{Fluctuations and Higher Moments of Conserved Charges from the Lattice}

\classification{\texttt{11.15.Ha, 12.38.Gc, 12.38.Mh, 25.75.-q}}

\author{P.~Hegde (for the HotQCD Collaboration)}{
  address={Physics Department, Brookhaven National Laboratory, Upton NY 11973, USA.}
}

\begin{abstract}
We present results for the lowest-order non-vanishing quark number susceptibilities. These were calculated using an improved action viz. the HISQ action, which controls taste violations that are responsible for distorting the light meson spectrum. Our calculations, with a pion mass of 160 MeV, are also much closer to the physical limit than previous studies. We find a broad crossover from the hadronic to the quark regimes, although interactions remain significant even at the highest temperatures studied. Our results are also in good agreement with Hadron Resonance Gas models below the crossover temperature.
\end{abstract}

\maketitle


The phase diagram of QCD at non-zero values of the quark chemical potentials $(\muq>0)$ is a subject of intense investigation. Progress is hampered by the fact that \emph{ab initio} numerical simulations (\emph{i.e.} the lattice) are currently not possible in this domain. On the other hand, from a variety of model studies it is believed that a second-order phase transition exists at some value of the baryochemical potential $\mu_B$. This picture is currently being tested in the low-energy runs at the Relativistic Heavy Ion Collider (RHIC) at Brookhaven~\cite{Aggarwal:2010wy, Aggarwal:2010pj}.

The basic observables in these experiments are the baryon and charge distributions, their fluctuations and higher moments. At vanishing chemical potential these moments may be calculated on the lattice by the method of Taylor expansions. These moments provide valuable information about the nature of the QCD chiral transition (which is a crossover for physical values of the light quark masses) as well as about the freezeout conditions. Furthermore, in the vicinity of the critical point these moments are expected to exhibit scaling behavior. Given sufficient statistics, such behavior is in principle observable and can serve as a signal for the elusive critical point.

\section{Second Order Susceptibilities}
The susceptibilities $\chi_2^i$ and $\chi_{11}^{ij}$ (where $i,j=u,d,s$) are defined as
\begin{align}
&&  \frac{\chi_2^i}     {T^2} = \frac{\partial^2}{\partial (\mu_i/T)^2} %
                                \left(\frac{P}{T^4}\right)\Bigg\vert_{\mu_i=0}
&& \text{and} 
&& \frac{\chi_{11}^{ij}}{T^2} = \frac{\partial^2}{\partial (\mu_i/T) \partial (\mu_j/T)} 
                                \left(\frac{P}{T^4}\right)\Bigg\vert_{\mu_i=\mu_j=0}. &&
\label{eq:taylor-exp}
\end{align}

$P$ and $T$ are the pressure and temperature, while the $\mu_i$'s are the quark chemical potentials. The diagonal susceptibilities $\chi_2^{u,d,s}$ measure fluctuations in the quark number distributions while the off-diagonal susceptibilities $\chi_{11}^{ij}$ measure the correlations between different flavors. By a straightforward change of basis, it is possible to obtain from these susceptibilities the corresponding ones in the baryon number $(B)$, electric charge $(Q)$ and strangeness $(S)$ basis. These latter susceptibilities are the ones measured in experiments.

Our calculations were performed with the staggered action. As is well-known, the staggered action describes four equivalent species (called tastes) of Dirac fermions in the continuum and thus the staggered spectrum contains sixteen pions instead of one. Taste symmetry is broken by interactions at finite lattice spacing and as a result fifteen of the pions become heavier by $\Od(\alpha_sa^2)$. This distortion of the light spectrum has consequences for thermodynamics, for e.g. the transition temperature is shifted upward. It also affects the calculation of the susceptibilities since these are sensitive to the light degrees of freedom. Fortunately a great deal of progress has been made toward designing actions that minimize these effects at a given lattice spacing. Currently the state-of-the-art in this respect is the HISQ action~\cite{Follana:2006rc}, which is also the action we used for our calculations.

We calculated these susceptibilities on $2+1$-flavor configurations at 10-15 temperatures in the range $125 < T < 340$ MeV. The temperature was varied by changing the coupling and at each coupling the quark masses were also tuned so as to stay on the line of constant physics {\it i.e.} the mass of the lightest Goldstone mode (identified with the pion) was fixed at approximately 160 MeV, while the mass of the corresponding strange meson (identified with the kaon) was fixed at the physical value of the kaon mass. The number of configurations used at each temperature varied from 500-700.

Our lattice sizes were $6\times 24^3$, $8\times 32^3$ and $12\times 48^3$. At any given temperature, a larger lattice implies a smaller spacing (which can be seen for e.g. via $T^{-1} = aN_\tau$). These numbers may be compared, for instance, with an earlier study by the RBC-Bielefeld Collaboration~\cite{Cheng:2008zh}. That study was done using lattices of size $4\times 16^3$ and $6\times 24^3$ {\it i.e.} farther from the continuum limit. Also, the calculation was done with $2+1$-flavor ensembles generated with the p4 action, which has greater taste symmetry breaking. Lastly, while the kaon mass was fixed at its physical value the pion was quite heavy, $m_\pi \approx 220$ MeV.

\begin{figure}[!thb]
\hspace{-0.07\textwidth}
\includegraphics[width=0.35\textwidth,height=0.20\textheight]{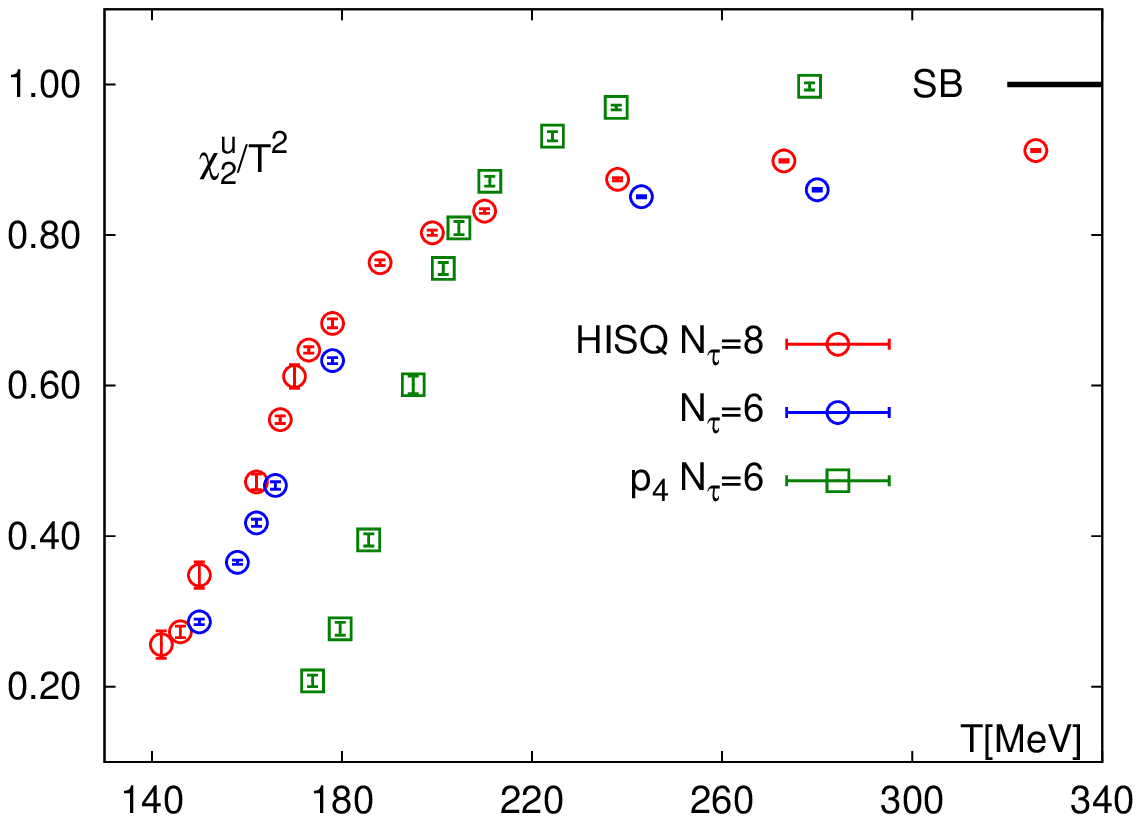}%
\hspace{0.17\textwidth}%
\includegraphics[width=0.35\textwidth,height=0.20\textheight]{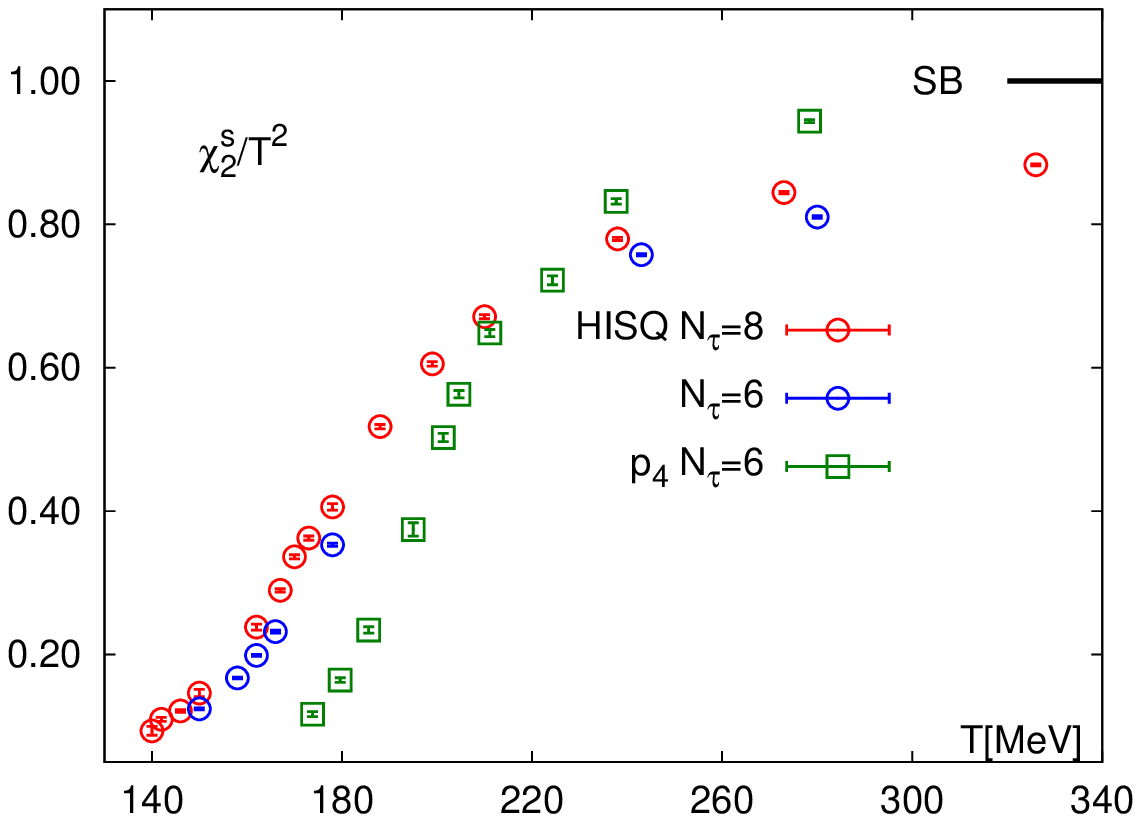}
\caption{Results for $\chi_2^u$ and $\chi_2^s$, compared with earlier results obtained using the p4 action~\cite{Cheng:2008zh}. The black lines marked ``SB'' denote the limiting value of an ideal gas of massless fermions.The transition temperature is $\approx190$ MeV for the p4 action at $N_\tau=6$, and $\approx160$ MeV at $N_\tau=8$ for the HISQ action.}
\label{fig:c2-hisq-p4}
\end{figure}

Fig.~\ref{fig:c2-hisq-p4} shows our results for the lowest-order susceptibilities $\chi_2^u$ and $\chi_2^s$. In comparing these with the earlier p4 results, two differences are immediately evident. On the one hand the crossover is smoother in the HISQ case, implying that the crossover from hadrons to quarks occurs over a broader range than was previously believed (This is also seen in the behavior of the off-diagonal susceptibilities $\chi_{BS}$ and $\chi_{QS}$). Furthermore the approach to the Stefan-Boltzmann limit is also slower and both the light and strange susceptibilities never attain this value even at the highest temperatures that we studied.

\section{Fluctuations of Conserved Charges and HRG}
\begin{figure}[!thb]
\hspace{-0.07\textwidth}
\includegraphics[width=0.35\textwidth,height=0.20\textheight]{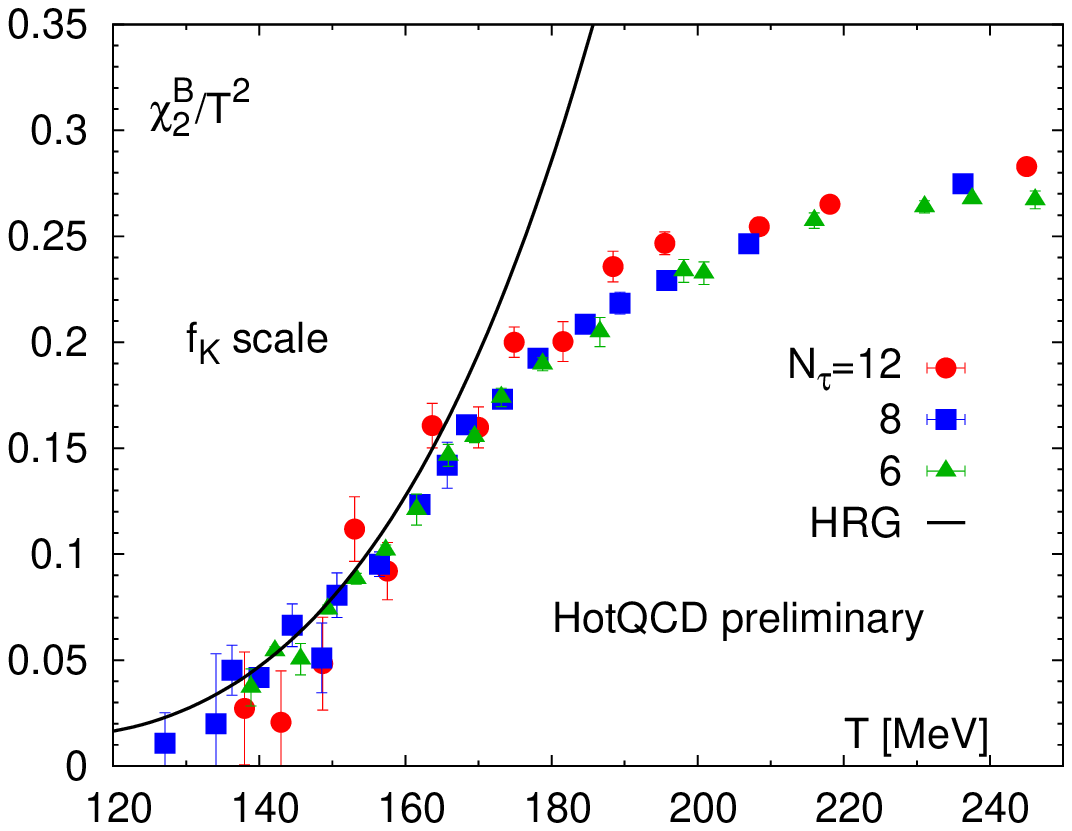}%
\hspace{0.17\textwidth}%
\includegraphics[width=0.35\textwidth,height=0.20\textheight]{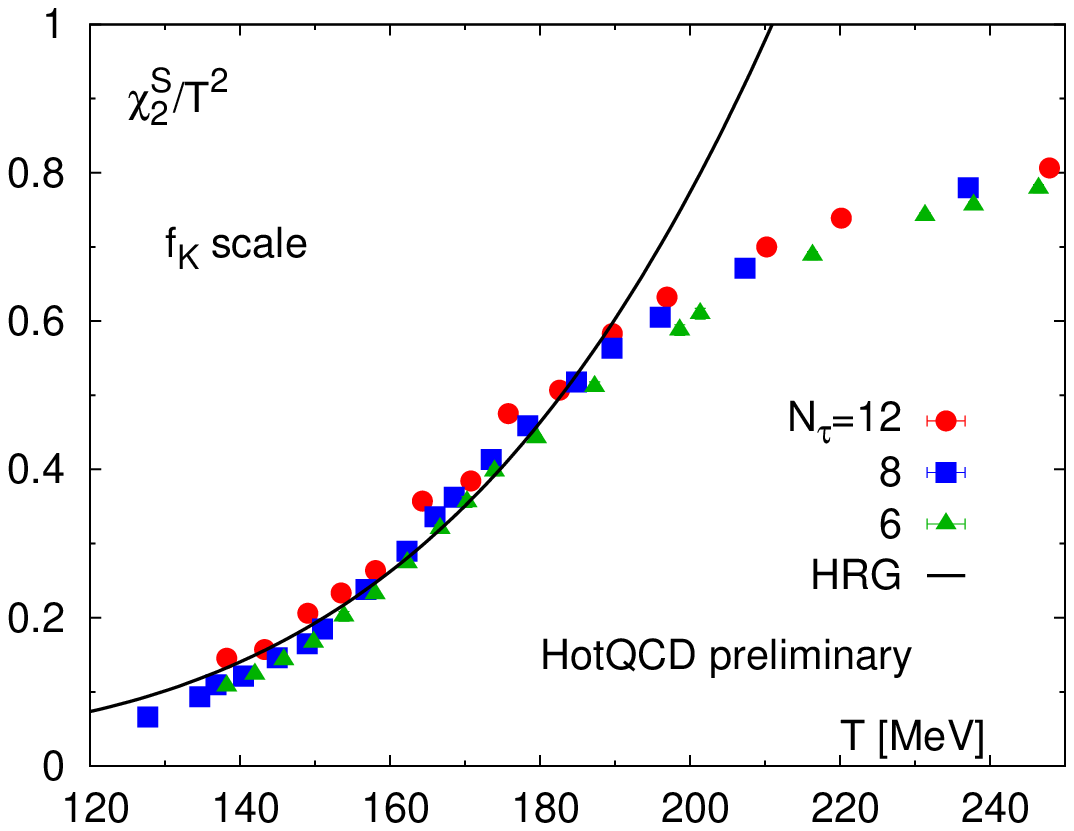}%
\caption{Baryon Number ($\chi_2^B$, left) and electric charge ($\chi_2^Q$, right) fluctuations. ``$f_K$ scale'' refers to the zero-temperature observable used to determine the temperature. The black curves are predictions from HRG models. All figures are preliminary~\cite{Bazavov:2011??}.}
\label{fig:c2BS-HRG}
\end{figure}

\begin{figure}[!thb]
\hspace{-0.07\textwidth}
\includegraphics[width=0.35\textwidth,height=0.20\textheight]{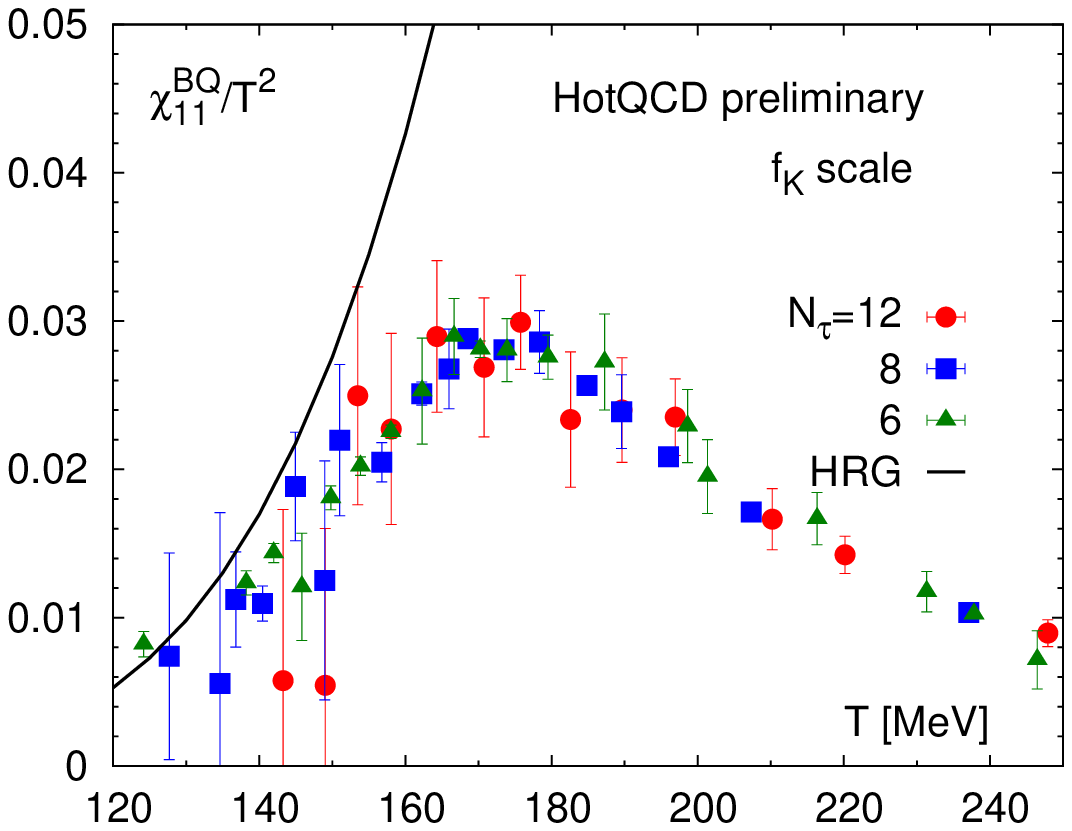}%
\hspace{0.17\textwidth}%
\includegraphics[width=0.35\textwidth,height=0.20\textheight]{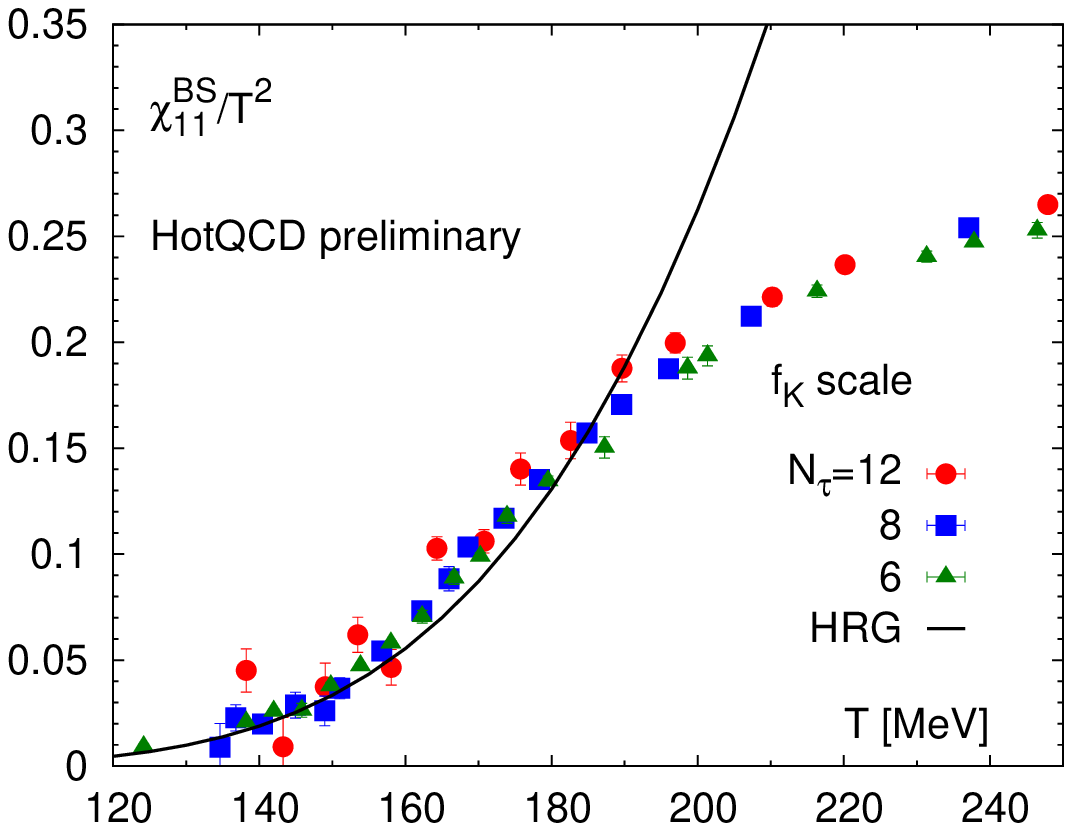}%
\caption{Off-diagonal correlations, between $B$ and $Q$ (left) and between $B$ and $S$ (right). The black curves are again predictions from HRG models, and ``$f_K$'' again refers to the observable used to determine the temperature. All figures are preliminary~\cite{Bazavov:2011??}.}
\label{fig:BQ-BS}
\end{figure}

The fluctuations of relevance to experiment are those of the baryon number, strangeness and electric charge. If we know all the $(u, d, s)$ fluctuations at a given order, we can compute the $(B,Q,S)$ fluctuations up to that order. Our results for the diagonal $B$ and $S$ fluctuations are shown in Fig.~\ref{fig:c2BS-HRG} while those for the off-diagonal susceptibilities are shown in Fig.~\ref{fig:BQ-BS}.

The solid lines in both plots are Hadron Resonance Gas (HRG) curves. HRG models have been found to provide a satisfactory description of nuclear matter at temperatures below the crossover~\cite{Karsch:2003zq, Allton:2005gk, Ratti:2010kj, Karsch:2010ck, Friman:2011pf}. The partition function is the product of those for ideal Bose (Fermi) gases, one for each meson (baryon) viz.
\begin{equation}
\ln\mathcal{Z} = \sum_{\pi,K,\eta,\dots}\ln\mathcal{Z}_\text{B-E}^\text{id.}(m_i,\mu_i) + \sum_{p,n,\Lambda,\dots}\ln\mathcal{Z}_\text{F-D}^\text{id.}(m_i,\mu_i).
\end{equation}
The low- and high-temperature limits of these susceptibilities have an especially simple interpretation. At very high temperatures the degrees of freedom will be quarks. Of these the $u$, $d$ and $s$ all carry baryon number but only the $s$ carries strangeness as well and therefore $\chi_{11}^{BS}\to 1/3$. Similarly, at low temperatures the lightest degrees of freedom carrying $B$ are the protons and neutrons, but only the protons carry both $B$ and $Q$. One therefore expects $\chi^{BQ}_{11}/\chi_2^B \to 1/2$ as $T\to0$.

To summarize, our calculations of the lowest quark and conserved charge susceptibilities have brought forth some interesting features. The transition from hadronic to QGP regimes was found to be broader than previously expected, and this was also reflected in correlations between the conserved charges. The Stefan-Boltzmann limit was not attained for $T\sim 2T_c$. On the hadronic side good agreement with HRG models was found with the latter providing both a qualitative understanding as well as a quantitative description of the data. In the future we shall look to calculate the higher-order susceptibilities. These are expected to exhibit scaling behavior and thus should be more sensitive to the nature of the phase transition.

\begin{theacknowledgments}
This work was supported in part by contract DE-AC02-98CH10886 of the U.S. Department of Energy (DoE).
\end  {theacknowledgments}

\bibliographystyle{aipproc}   

\end{document}